\documentclass[pdftex]{article}
\usepackage{icrctc07}

\pdfoutput=1 

\def\degr{\hbox{$^\circ$}}

\title{Morphological Studies of the PWN Candidate HESS J1809-193}
\shorttitle{The PWN candidate HESS J1809-193}
\authors{Nu. Komin$^{1}$,
  S. Carrigan$^{2}$,
   A. Djannati-Ata\"i$^{3}$,
   Y.A. Gallant$^{1}$,\\
   K. Kosack$^{2}$,
   G. Puehlhofer$^{4}$,
   S. Schwemmer$^{4}$,
   for the H.E.S.S. Collaboration$^{5}$}
\shortauthors{Komin and et al}
\afiliations{$^1$ LPTA, Universit\'e Montpellier 2, CNRS/IN2P3, Montpellier, France\\ 
$^2$ Max-Planck-Institut f\"ur Kernphysik, Heidelberg, Germany\\
$^3$ APC (CNRS, Universit\'e Paris VII, CEA, Observatoire de Paris), Paris, France\\
$^4$ Landessternwarte, Universit\"at Heidelberg, Germany\\
$^5$ {\tt www.mpi-hd.mpg.de/HESS }}
\email{komin@lpta.in2p3.fr}

\abstract{The source HESS J1809$-$193 was discovered in 2006 in data
of the Galactic Plane survey, followed by several re-observations. It
shows a hard gamma-ray spectrum and the emission is clearly
extended. Its vicinity to PSR\,J1809-1917, a high spin-down luminosity
pulsar powerful enough to drive the observed gamma-ray emission, makes
it a plausible candidate for a TeV Pulsar Wind Nebula (PWN). On the
other hand, in this region of the sky a number of faint,
radio-emitting supernova remnants can be found, making a firm
conclusion on the source type difficult.

Here we present a detailed morphological study of recent H.E.S.S. data
and compare the result with X-ray measurements taken with
\emph{Chandra} and radio data. The association with a PWN is likely,
but contributions from supernova remnants cannot be ruled out.

}

\begin{document}
\maketitle

\section{Introduction}

Since the beginning of observations with H.E.S.S. (High Energy
Stereoscopic System) in 2003 the number of known TeV gamma-ray
emitting sources has increased drastically. The ongoing scan of the
Galactic plane revealed several bright and extended sources for which
no clear association with objects in other wavelength could be found
\cite{Scan1,Scan2}.

Pulsars, rapidly rotating neutron stars, are widely believed to be
able to accelerate particles up to PeV energies. Those objects loose
their rotational energy in winds of relativistic particles.  The
confinement of the wind in the interaction with the ambient
interstellar material forms shocks; the particles accelerated there
are visible as a Pulsar Wind Nebula (PWN) (see \cite{GaenslerSlane}
for a review). Synchrotron radiation seen in radio and X-rays prove
the existence of relativistic electrons in the PWN. These electrons
undergo inverse Compton (IC) scattering off ambient radiation fields,
like the Cosmic Microwave Background, Galactic infrared background and
optical star light, leading to the production of TeV gamma-rays.

Here we present the observation of one TeV source, HESS\,J1809$-$193,
which is located close to a powerful pulsar and thus a good PWN
candidate. X-ray emission from the direction of the pulsar support the
theory of being a PWN. However, confusion with other sources cannot be
ruled out.

\section{TeV observations of  HESS\,J1809$-$193}

H.E.S.S. is a system of four Imaging Atmospheric Cherenkov telescopes
(IACTs) dedicated to the observation of TeV gamma-rays. Its high
sensitivity allows the detection of point sources with a flux of $1\%$
of that of the Crab nebula within 25\,h \cite{Crab}. Its large field
of view and an angular resolution of better than $0.1\degr$ makes it
an ideal tool for observations of extended objects and for the
conduction of sky surveys.

In the original Galactic plane survey conducted with H.E.S.S., TeV
emission from HESS\,J1809$-$193 was only marginally detected. Further
re-observations confirmed the existence of gamma-ray
emission \cite{paper}. Further observations were performed in autumn
2006; in total data with a live time of 32\,h is available.

\begin{figure}
\begin{center}
\includegraphics [width=0.48\textwidth]{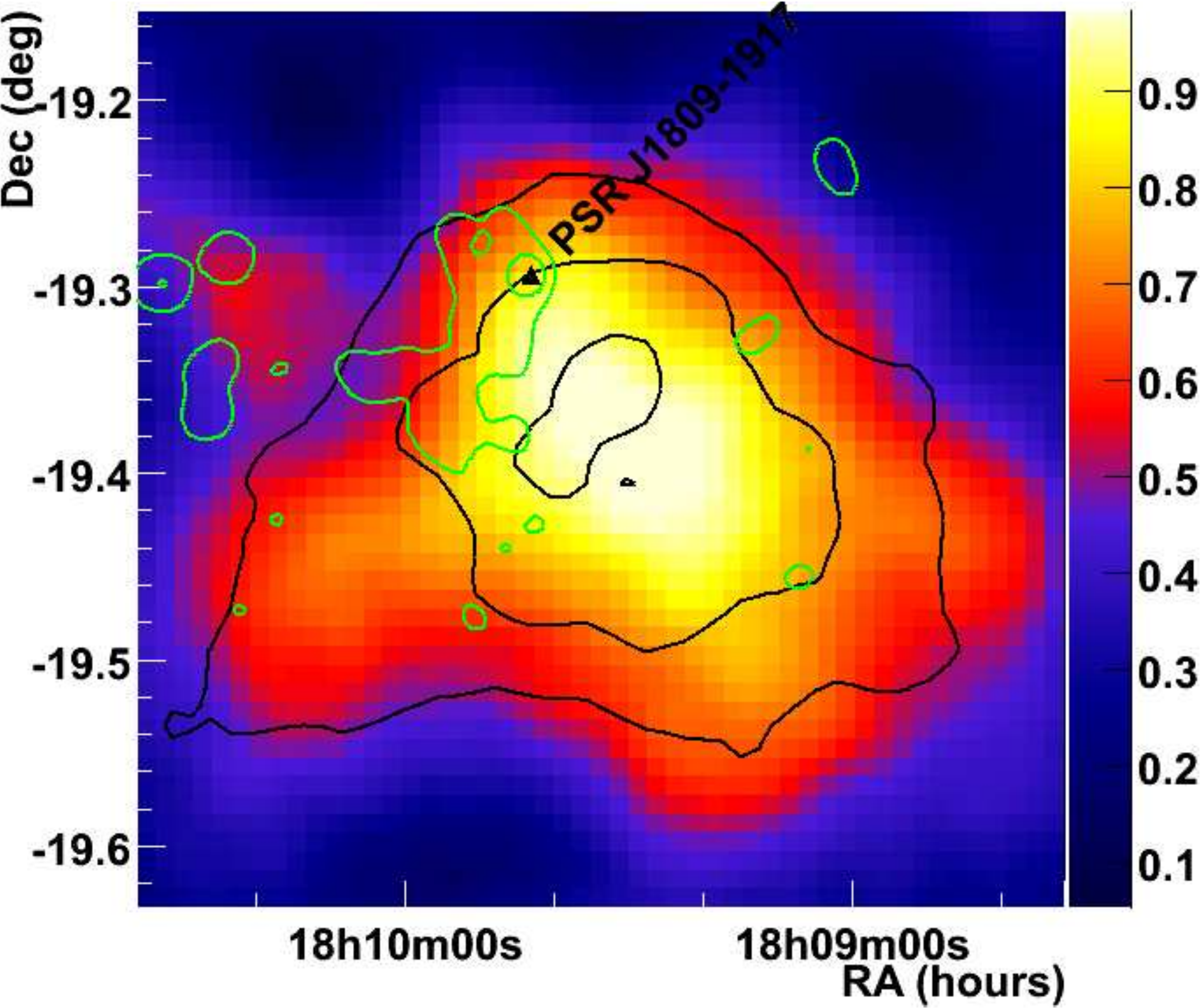}
\end{center}
\caption{TeV gamma-ray excess counts from the direction of
HESS\,J1809$-$193 (colour scale). The image is smoothed with the
point-spread function. Overlaid are 5, 7 and 9\,$\sigma$ significance
contours, oversampled with a circle with radius $0.1\degr$. The
position of the pulsar is marked with a black triangle. The green
contours denote the diffuse X-ray emission from
Fig.~\ref{fig:chandra}.}
\label{fig:skymap}
\end{figure}

The gamma-ray excess map of the source HESS\,J1809$-$193 is shown in
Fig.~\ref{fig:skymap}. The emission is clearly extended lying
south-west of the pulsar. In addition faint emission can be seen to
the south-east. In total, an excess of 3600 events with a significance
of $19\,\sigma$ was detected. The source shows an energy spectrum
consistent with a power law with an index of $2.2 \pm
0.1_\mathrm{stat} \pm 0.2_\mathrm{syst}$ and an energy flux between 1
and 10\,TeV of roughly $1.3 \times 10^{-11} \,
\mathrm{erg\,cm}^{-2}\,\mathrm{s}^{-1}$ \cite{paper}. If this energy
flux is projected to the distance of the pulsar, only $1.2\%$ of the
pulsar's spin down luminosity of
$1.8\times10^{36}\,\mathrm{erg\,s}^{-1}$ is needed to power the
H.E.S.S. source. Therefore it seems to be plausible that
HESS~J1809$-$193 is indeed a Pulsar Wind Nebula.

\section{X-Ray Observations}

In the data of the Galactic plane scan performed with the ASCA
satellite diffuse emission was detected \cite{Bamba}, which turned out
to be coincident with the TeV source. The X-ray source G$11.0+0.0$ has
been discussed to be either a young shell-type supernova remnant (SNR)
or a plerionic SNR.

\begin{figure}
\begin{center}
\includegraphics [width=0.48\textwidth]{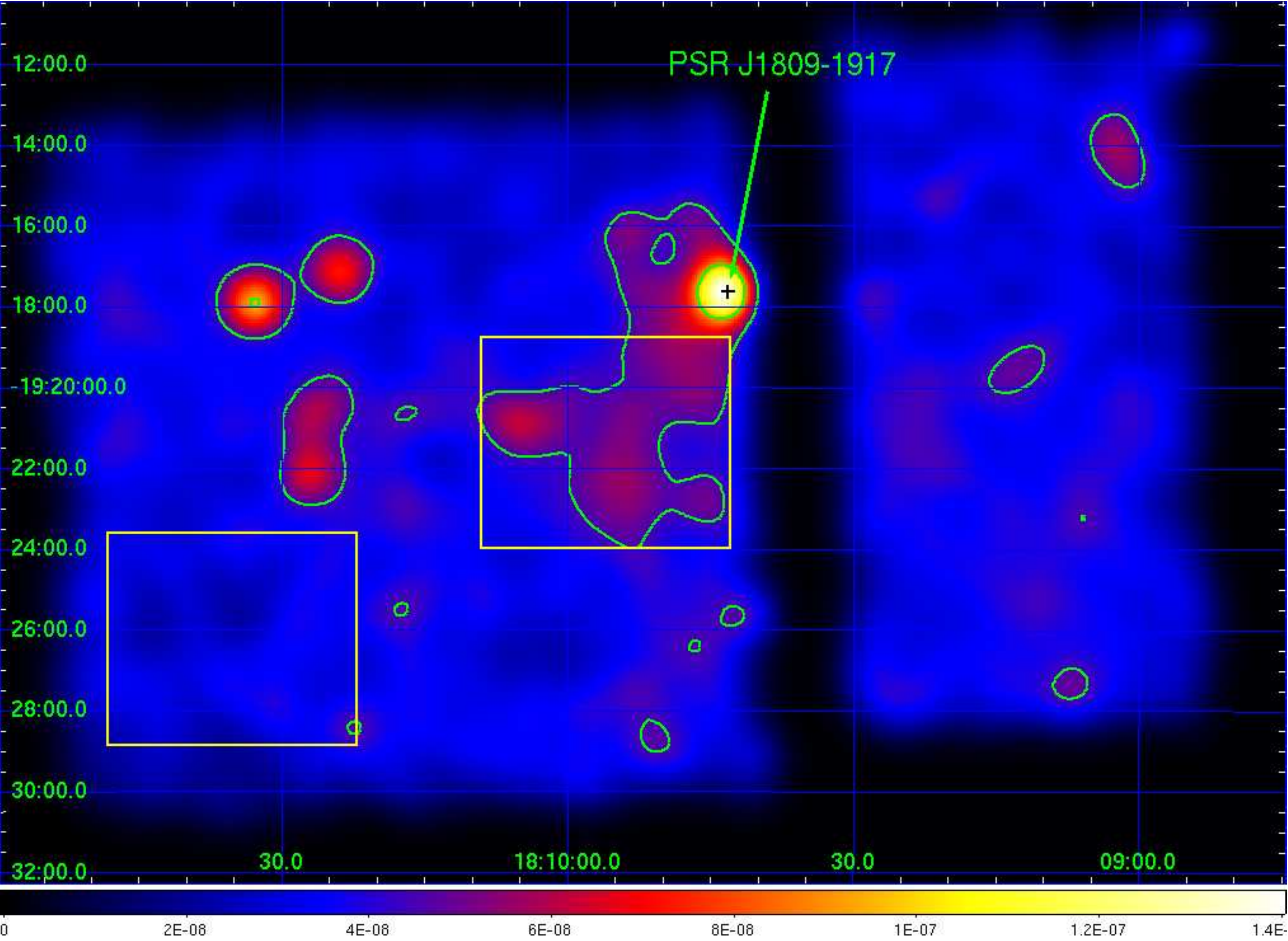}
\end{center}
\caption{Chandra X-ray excess of the field of view of HESS
  J1809$-$193. The map is exposure-corrected and smoothed with a
  Gaussian with $32''$.}
\label{fig:chandra}
\end{figure}

High-resolution observations with the \emph{Chandra} satellite revealed
a compact X-ray nebula north of the pulsar and additional faint
emission south \cite{PSR}. Here we present \emph{Chandra} data which
was taken in February 2007 (ObsID 6720).  Figure~\ref{fig:chandra}
shows the exposure-corrected and smoothed X-ray excess map. It shows a
strong X-ray nebula, high resolution images show its extension to the
north of the pulsar \cite{PSR}. Further faint emission can be seen to
the south.

The significance of the diffuse emission was tested by comparing the
on-source region with an off-source region in the same field of view
(these regions are indicated by the yellow rectangles in
Fig.~\ref{fig:chandra}). Taking into account the small acceptance
difference of 4\% (estimated from the exposure map at 2 keV), the
source region shows an excess of about 900 events with a statistical
significance of $10\,\sigma$.

The contour of the diffuse X-ray emission is overlaid in the TeV
excess map in Fig.~\ref{fig:skymap}. It should be noted that due to
the gap between the chips of the X-ray detector and the different
nature of the chips on the right hand side, no conclusions can be
drawn on the existence of diffuse emission to the west. However, it
can be seen that the X-ray nebula's extension to the south is far
smaller than the extension of the TeV emission.

\section{Radio Data of the Field of View}

\begin{figure}
\begin{center}
\includegraphics [width=0.48\textwidth]{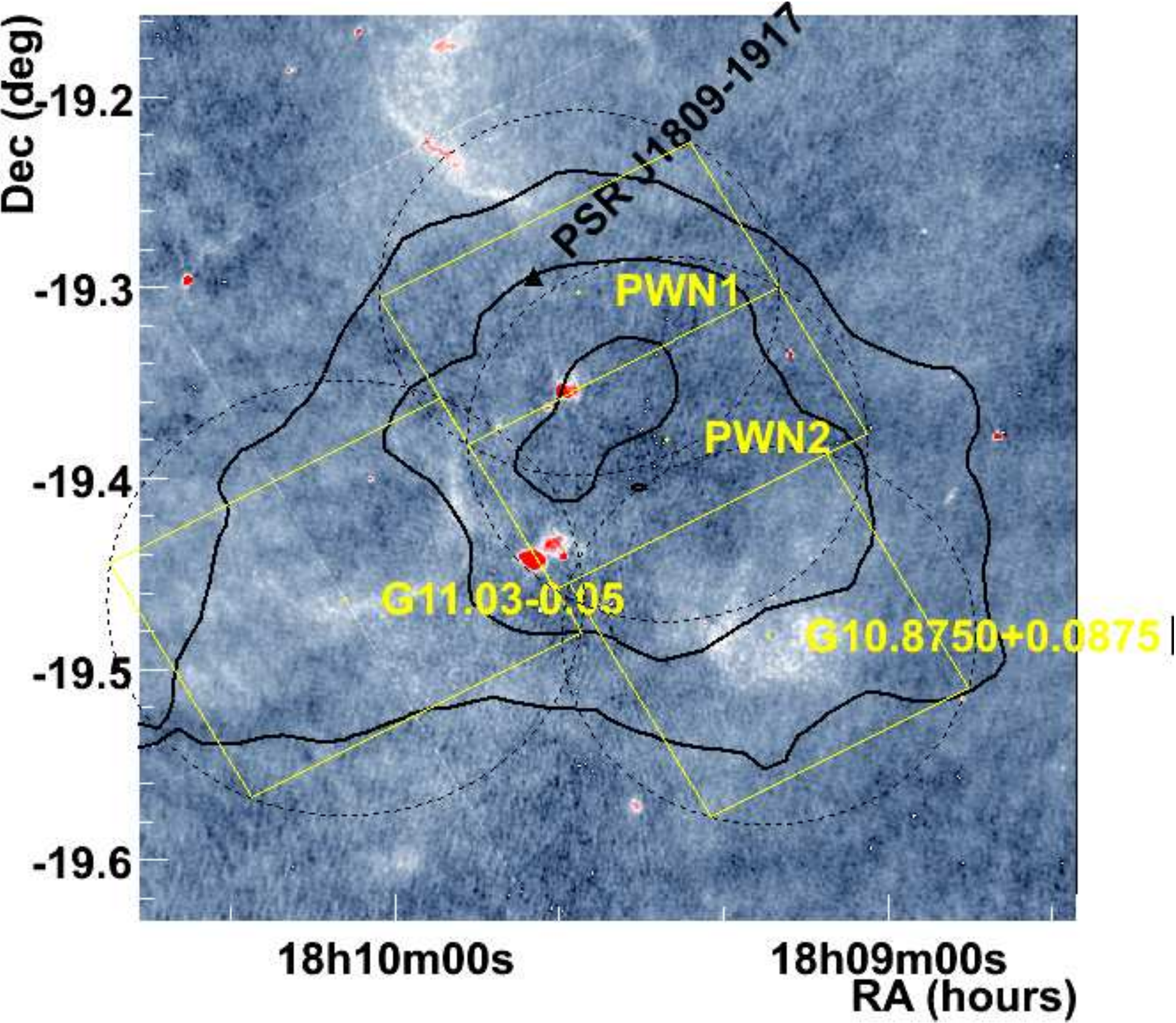}
\end{center}
\caption{MAGPIS radio image. Overlaid are the 5, 7 and 9\,$\sigma$
  H.E.S.S. significance contours. The yellow rectangles indicate the
  test regions for the spectral analysis of the H.E.S.S. data.}
\label{fig:radio}
\end{figure}

We compared the region of HESS~J1809$-$193 with radio data of the
Multi-Array Galactic Plane Imaging Survey \cite{MAGPIS} shown in
Fig.~\ref{fig:radio}. North of the pulsar is the SNR G$11.18+0.11$
\cite{Brogan,Green}, not coincident with the H.E.S.S. excess. Located
south-east of the pulsar and coincident with the H.E.S.S. excess is
the SNR G$11.03-0.05$ \cite{Brogan,Green}. Further south-west of the
pulsar is the supernova remnant candidate $10.8750+0.0875$
\cite{MAGPIS}. In the region between the latter two SNRs and the
pulsar no diffuse radio emission can be found.

For a spectral analysis different regions according to the radio data
have been chosen. Two regions for the SNRs G$11.03-0.05$ and
$10.8750+0.0875$ and another two regions for the possible pulsar wind
nebula (PWN1, PWN2). These regions are indicated in
Fig.~\ref{fig:radio}.

\section{Spectral Analysis}

\begin{figure}
\begin{center}
\includegraphics [width=0.48\textwidth]{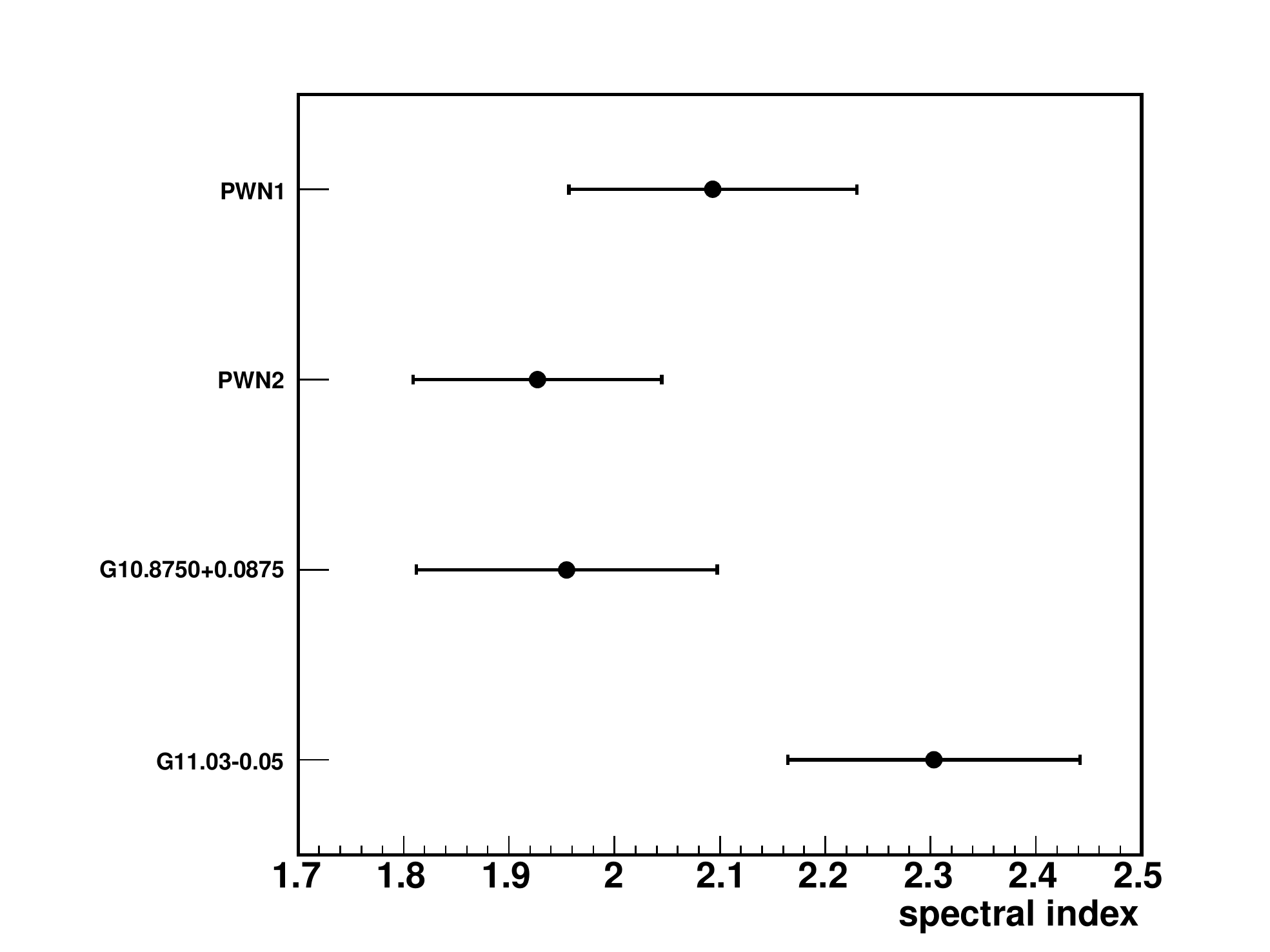}
\end{center}
\caption{Spectral indices for the test regions indicated in
Fig.~\ref{fig:radio}.}
\label{fig:spectrum}
\end{figure}

The TeV energy spectra of each region were fitted with a power
law. The spectral indices are shown in Fig.~\ref{fig:spectrum}.  The
regions PWN1, PWN2 and $10.8750+0.0875$ are with increasing distance
to the pulsar. A steepening of the spectrum with increasing distance
would be a clear indication for a PWN (see \cite{J1825_2}). A
significant different energy spectrum between the region of the SNR
G$11.03-0.05$ and the rest of the emission region would be a hint for
different source associations. However, due to the large statistic
uncertainties no conclusion on spectral variations over the extension
of the source can be drawn.

\section{Discussion}

The existence of the powerful pulsar PSR~J$1809-1917$ which can easily
power the TeV emission suggests that the TeV emission is a PWN
associated with the pulsar. The association with a PWN is further
supported by a compact X-ray nebula and diffuse X-ray emission
coincident with the H.E.S.S. source. The X-ray emission is
significantly smaller than the TeV source. This has been already seen
for the PWN HESS~J$1825-137$ \cite{J1825}.

On the other hand, two supernova remnants coincide with the TeV
emission. They do not show X-ray emission, however, TeV emission can
still be expected, in particular if if they are associated with dense
molecular clouds \cite{W28,Yamazaki}. Further studies will include the
search for dense molecular clouds in the region.

\section{Conclusion}

Detailed studies of the source HESS~J$1809-193$ and comparison with
objects in other wavelength show that this source is likely a PWN
powered by the pulsar PSR~J$1809-1917$.  The number of TeV emitting
PWNe is increasing, showing that TeV PWN constitute a significant
fraction of the Galactic TeV gamma-ray source population.

Contribution of gamma-ray emission from faint radio supernova remnants
cannot be ruled out. Radio-emitting, X-ray quiet SNRs, possibly in
connection with dense molecular clouds, remain interesting targets for
gamma-ray observations.

\section{Acknowledgements}

The support of the Namibian authorities and of the University of
Namibia in facilitating the construction and operation of H.E.S.S. is
gratefully acknowledged, as is the support by the German Ministry for
Education and Research (BMBF), the Max Planck Society, the French
Ministry for Research, the CNRS-IN2P3 and the Astroparticle
Interdisciplinary Programme of the CNRS, the U.K. Science and
Technology Facilities Council (STFC), the IPNP of the Charles
University, the Polish Ministry of Science and Higher Education, the
South African Department of Science and Technology and National
Research Foundation, and by the University of Namibia. We appreciate
the excellent work of the technical support staff in Berlin, Durham,
Hamburg, Heidelberg, Palaiseau, Paris, Saclay, and in Namibia in the
construction and operation of the equipment.

\bibliography{icrc1036}
\bibliographystyle{plain}

\end{document}